\documentclass[twoside,twocolumn,9pt]{article}
\usepackage{extsizes}
\usepackage[super,sort&compress,comma]{natbib} 
\usepackage[version=3]{mhchem}
\usepackage[left=1.5cm, right=1.5cm, top=1.785cm, bottom=2.0cm]{geometry}
\usepackage{balance}
\usepackage{times,mathptmx}
\usepackage{sectsty}
\usepackage{graphicx} 
\usepackage{lastpage}
\usepackage[format=plain,justification=justified,singlelinecheck=false,font={stretch=1.125,small,sf},labelfont=bf,labelsep=space]{caption}
\usepackage{float}
\usepackage{fancyhdr}
\usepackage{fnpos}
\usepackage[english]{babel}
\addto{\captionsenglish}{%
  
}
\usepackage{array}
\usepackage{droidsans}
\usepackage{charter}
\usepackage[T1]{fontenc}
\usepackage[usenames,dvipsnames]{xcolor}
\usepackage{setspace}
\usepackage[compact]{titlesec}
\usepackage{hyperref}
\usepackage{epstopdf}%This line makes .eps figures into .pdf - please comment out if not required.
\usepackage{amsmath,amssymb}

\definecolor{cream}{RGB}{222,217,201}
\DeclareMathOperator{\sign}{sign}

\begin{document}

\pagestyle{fancy}
\thispagestyle{plain}
\fancypagestyle{plain}{

%%%HEADER%%%
\fancyhead[C]{\includegraphics[width=18.5cm]{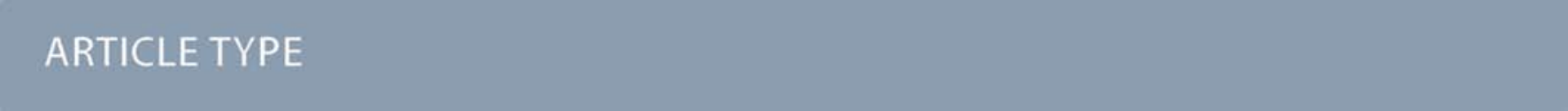}}
\fancyhead[L]{\hspace{0cm}\vspace{1.5cm}\includegraphics[height=30pt]{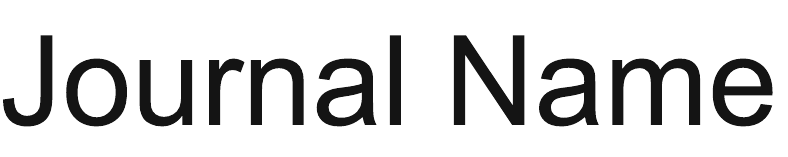}}
\fancyhead[R]{\hspace{0cm}\vspace{1.7cm}\includegraphics[height=55pt]{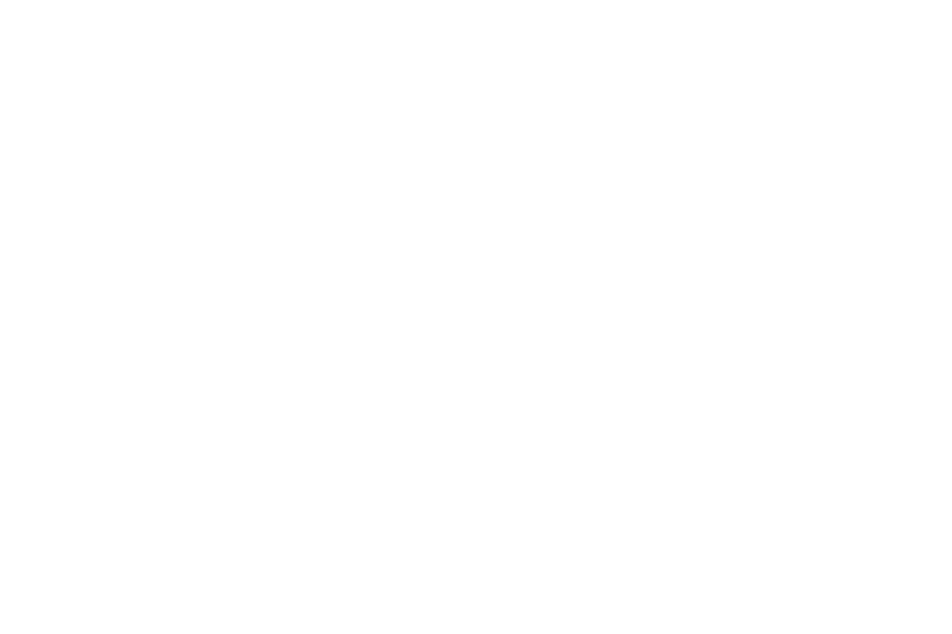}}
\renewcommand{\headrulewidth}{0pt}
}
%%%END OF HEADER%%%

%%%PAGE SETUP - Please do not change any commands within this section%%%
\makeFNbottom
\makeatletter
\renewcommand\LARGE{\@setfontsize\LARGE{15pt}{17}}
\renewcommand\Large{\@setfontsize\Large{12pt}{14}}
\renewcommand\large{\@setfontsize\large{10pt}{12}}
\renewcommand\footnotesize{\@setfontsize\footnotesize{7pt}{10}}
\renewcommand\scriptsize{\@setfontsize\scriptsize{7pt}{7}}
\makeatother

\renewcommand{\thefootnote}{\fnsymbol{footnote}}
\renewcommand\footnoterule{\vspace*{1pt}% 
\color{cream}\hrule width 3.5in height 0.4pt \color{black} \vspace*{5pt}} 
\setcounter{secnumdepth}{5}

\makeatletter 
\renewcommand\@biblabel[1]{#1}            
\renewcommand\@makefntext[1]% 
{\noindent\makebox[0pt][r]{\@thefnmark\,}#1}
\makeatother 
\renewcommand{\figurename}{\small{Fig.}~}
\sectionfont{\sffamily\Large}
\subsectionfont{\normalsize}
\subsubsectionfont{\bf}
\setstretch{1.125} %In particular, please do not alter this line.
\setlength{\skip\footins}{0.8cm}
\setlength{\footnotesep}{0.25cm}
\setlength{\jot}{10pt}
\titlespacing*{\section}{0pt}{4pt}{4pt}
\titlespacing*{\subsection}{0pt}{15pt}{1pt}
%%%END OF PAGE SETUP%%%

%%%FOOTER%%%
\fancyfoot{}
\fancyfoot[LO,RE]{\vspace{-7.1pt}\includegraphics[height=9pt]{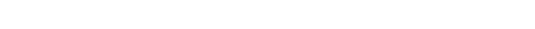}}
\fancyfoot[CO]{\vspace{-7.1pt}\hspace{13.2cm}\includegraphics{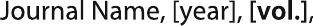}}
\fancyfoot[CE]{\vspace{-7.2pt}\hspace{-14.2cm}\includegraphics{RF}}
\fancyfoot[RO]{\footnotesize{\sffamily{1--\pageref{LastPage} ~\textbar  \hspace{2pt}\thepage}}}
\fancyfoot[LE]{\footnotesize{\sffamily{\thepage~\textbar\hspace{3.45cm} 1--\pageref{LastPage}}}}
\fancyhead{}
\renewcommand{\headrulewidth}{0pt} 
\renewcommand{\footrulewidth}{0pt}
\setlength{\arrayrulewidth}{1pt}
\setlength{\columnsep}{6.5mm}
\setlength\bibsep{1pt}
%%%END OF FOOTER%%%

%%%FIGURE SETUP - please do not change any commands within this section%%%
\makeatletter 
\newlength{\figrulesep} 
\setlength{\figrulesep}{0.5\textfloatsep} 

\newcommand{\topfigrule}{\vspace*{-1pt}% 
\noindent{\color{cream}\rule[-\figrulesep]{\columnwidth}{1.5pt}} }

\newcommand{\botfigrule}{\vspace*{-2pt}% 
\noindent{\color{cream}\rule[\figrulesep]{\columnwidth}{1.5pt}} }

\newcommand{\dblfigrule}{\vspace*{-1pt}% 
\noindent{\color{cream}\rule[-\figrulesep]{\textwidth}{1.5pt}} }

\makeatother
%%%END OF FIGURE SETUP%%%

%%%TITLE AND AUTHORS%%%
\twocolumn[
  \begin{@twocolumnfalse}
\vspace{3cm}
\sffamily
\begin{tabular}{m{4.5cm} p{13.5cm} }

\includegraphics{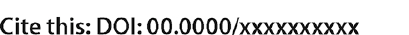} & \noindent\LARGE{\textbf{Dynamics of individual Brownian rods  in a microchannel flow}} \\%Article title goes here instead of the text "This is the title"
 & \vspace{0.3cm} \\

 & \noindent\large{Andreas Z\"{o}ttl,$^{\ast}$\textit{$^{ab}$} Kira E. Klop,\textit{$^{c}$} Andrew K. Balin,\textit{$^{b}$} Yongxiang Gao,\textit{$^{c,d}$} Julia M. Yeomans,\textit{$^{b}$}  and Dirk G. A. L. Aarts\textit{$^{c}$}} \\%Author names go here instead of "Full name", etc.

\includegraphics{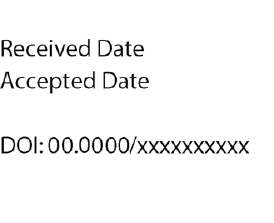} & \\

\end{tabular}

 \end{@twocolumnfalse} \vspace{0.6cm}

 ] 
%%%END OF TITLE AND AUTHORS%%%

%%%FONT SETUP - please do not change any commands within this section
\renewcommand*\rmdefault{bch}\normalfont\upshape
\rmfamily
\section*{}
\vspace{-1cm}

%%%FOOTNOTES%%%

\footnotetext{
  \textit{$^{a}$~Institute for Theoretical Physics, TU Wien, Wiedner Hauptstra{\ss}e 8-10, Wien, Austria.  E-mail: andreas.zoettl@tuwien.ac.at} \\
  \textit{$^{b}$~The Rudolf Peierls Centre for Theoretical Physics, University of Oxford, Clarendon Lab., Parks Rd., Oxford, OX1 3PU, UK.} \\
  \textit{$^{c}$~Department of Chemistry, Physical and Theoretical Chemistry Laboratory, University of Oxford, Oxford, OX1 3QZ, UK.} \\
  \textit{$^{d}$~Institute for Advanced Study, Shenzhen University, Nanhai Avenue 3688, Nanshan District, Shenzhen, 518060, China.}
}

%Please use \dag to cite the ESI in the main text of the article.
%If you article does not have ESI please remove the the \dag symbol from the title and the footnotetext below.
%%%%%\footnotetext{\dag~Electronic Supplementary Information (ESI) available: [details of any supplementary information available should be included here]. See DOI: 00.0000/00000000.}
%additional addresses can be cited as above using the lower-case letters, c, d, e... If all authors are from the same address, no letter is required

%\footnotetext{\ddag~These authors contributed equally to this work.}

%%%END OF FOOTNOTES%%%

%%%ABSTRACT%%%%

\sffamily{\textbf{We study the orientational dynamics of heavy silica microrods flowing through a microfluidic channel. Comparing experiments and Brownian dynamics simulations we identify different particle orbits, in particular in-plane tumbling behavior, which cannot be explained by classical Jeffery theory, and we relate this behavior to the rotational diffusion of the rods. By constructing the full, three-dimensional, orientation distribution, we describe the rod  trajectories and quantify the persistence of Jeffery orbits using temporal correlation functions of the Jeffery constant. We find that our colloidal rods lose memory of their initial configuration in about a second, corresponding to half a Jeffery period. }}\\%The abstrast goes here instead of the text "The abstract should be..."

%%%END OF ABSTRACT%%%%

\rmfamily %Please do not remove this line.

%%%MAIN TEXT%%%%

\begin{figure}
%\centering
\includegraphics[width=\columnwidth]{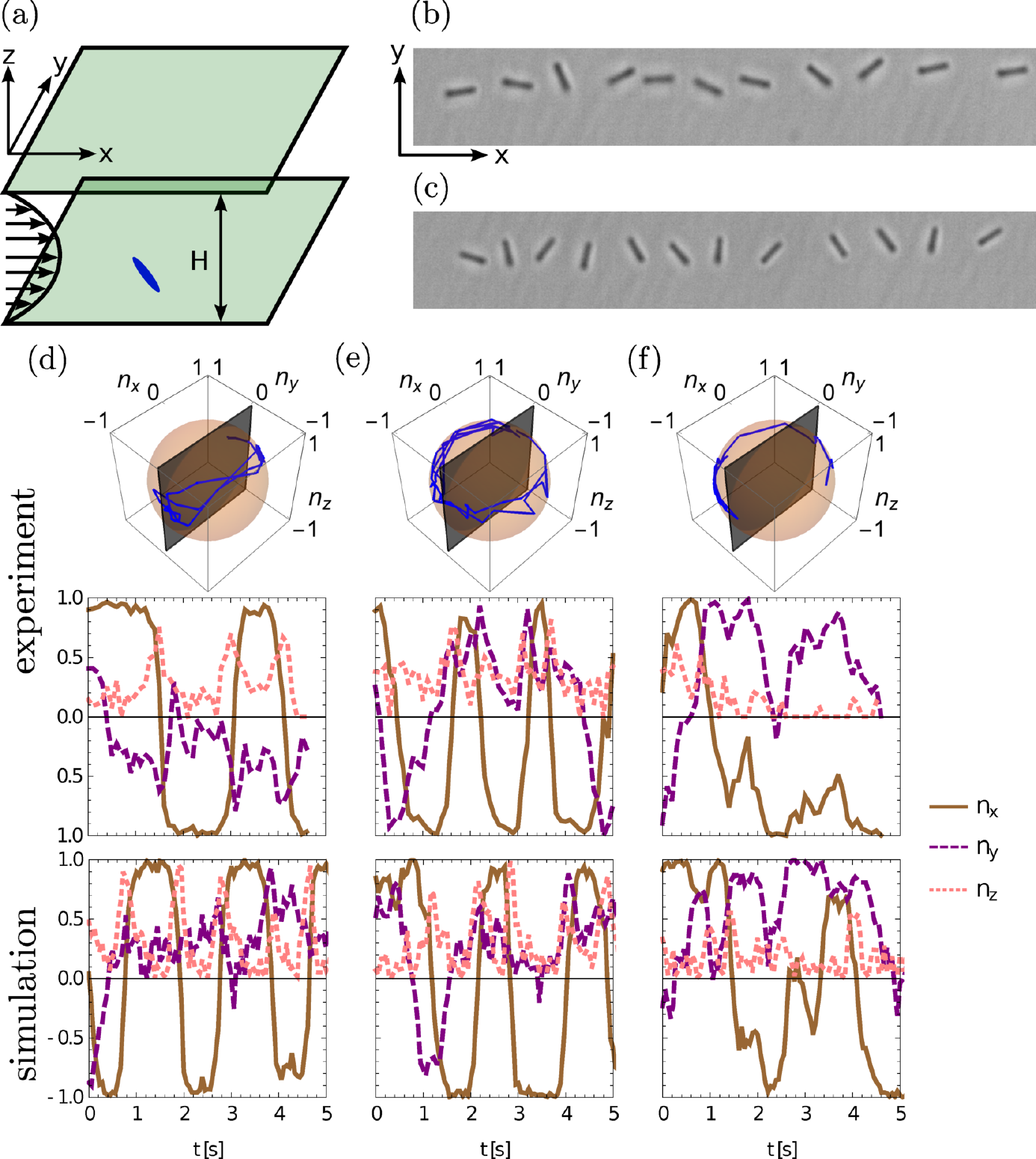}
\caption{ %\ju{rewritten}
(a) Channel geometry. The size of the rod and the channel height $H$ are drawn to scale.
(b) and (c) show a single rod at different times as it moves through the channel, with $0.3s$ between each image.
(b) Typical $xy$ tumbling. %\juc{would it be better to invert these, as it looks like the rods are at the top of the channel}
(c) Typical kayaking-like motion.
(d-f) rod trajectories for  kayaking (d), $xy$ tumbling (e) and more irregular motion (f). In each case the top frame shows motion on the unit sphere from experiment. The middle frame shows 
reconstructed experimental rod orientations $n_x$, $n_y$, $n_z$. The bottom panel shows $n_x$, $n_y$, $n_z$  for similar classes of orbits 
obtained from Brownian dynamics simulations. %\ju{if we go for this caption will need to remove g,h,i}
}
%(c) and (d)  plot of the $x$, $y$ and $z$ components of the orientation.
 \label{Fig:1}
%\juc{I'd show Jeffrey tunbling, Jeffrey kayaking, xy tumbling, random to make it easier to explain}}

\end{figure}

Understanding the physics of a micron-scale, elongated particle moving in viscous flows is widely relevant.
Examples include the effect of shape
on nano-particle assisted drug delivery \cite{BS15}, the dynamics of flowing suspensions of bacteria \cite{JF19}, nanoengineering optically anisotropic devices \cite{FL05} and the processing of food and pastes \cite{D15}. 
The orientational behaviour of a single, non-Brownian,  ellipsoidal rod in a simple shear flow was analysed theoretically in a classic paper by Jeffery in 1922 \cite{Jeffery} who found that the orientation of an axisymmetric ellipsoid undergoes a periodic motion on the unit sphere.
%The orientation of axisymmetric particles can be represented by a unit vector $\mathbf{n}=n_x\hat{\mathbf{x}}+n_y\hat{\mathbf{y}}+n_z\hat{\mathbf{z}}$ \az{pointing along the long axis of the particle}.
%Jeffery \cite{Jeffery} showed that the vector $\mathbf{n}$ of non-Brownian axisymmetric ellipsoidal particles in a linear shear flow performs periodic motion on a unit sphere. 
Assuming that the flow is in the $x$ direction and that the shear gradient is along $z$ [see also Fig.~\ref{Fig:1}(a) for non-uniform shear],
the specific orbit a particle follows is determined by its aspect ratio $\lambda$, the shear rate $\dot{\gamma}$ and the Jeffery constant 
$C=\sqrt{n_x^2+n_z^2/\lambda^2}/n_y$ 
which depends on the initial orientation of the rod and ranges from -$\infty$ to $\infty$. Here, the unit vector $\mathbf{n}=n_x\hat{\mathbf{x}}+n_y\hat{\mathbf{y}}+n_z\hat{\mathbf{z}}$ points along the long axis of the particle.
For $C = \pm \infty$ %the orbit is a circular motion in the $xz$-plane. 
the rod rotates in the $xz$-plane. It is oriented along the direction of flow 
%\juc{question - I thought it was at an angle to this because of he shear} 
for most of the time, but periodically ``tumbles'', i.e.\  flips its orientation by $180^\circ$.
% This type of motion is called ``tumbling'' in the literature. 
For smaller values of $C$, the motion has a finite $y$-component, and looks very similar to the trajectory of the paddles of a kayak if viewed from the side, hence the term ``kayaking'' for these orbits.
At $C=0$, the rod  orients along the $y$-direction and only rotates around its long axis. This last type of motion is called ``log-rolling'', and has been shown to be unstable for rod-shaped particles \cite{logrolling}.
Experiments on single rods, carried out in the non-Brownian regime, have demonstrated the applicability of Jeffery's ideas \cite{Einarsson}.

  Subsequent research has shown how the many perturbations present in flowing channels can affect the reproducibility and longevity of the orbits. For example, even small deviations from a perfect axisymmetric rod shape can lead to the appearance of doubly periodic and chaotic orbits, and these have been studied both theoretically and experimentally \cite{HinchLeal,Yarin1997,Mehlig2016}. The proximity of channel walls \cite{HZ01,ZB07,AK13,HA16,MJ18}, inertia \cite{logrolling} and the viscoelasticity of the shearing fluid \cite{GS08} have also been shown to perturb the Jeffery solution. Furthermore, noise may also affect the orbits: 
  %  Jeffery's theory is in excellent agreement with experiments
  %\juc{exact?}
%  for  axisymmetric particles in a simple shear, where noise does not play a role.
for smaller rods, where Brownian motion (i.e.\ thermal noise) is relevant, there has been work using rheo-optical techniques to measure rod distributions \cite{FF86,XL15}, and these have been been compared to theoretical predictions and Brownian dynamics simulations \cite{HL73,HZ02,LK15,PO18}. However, we are not aware of any existing experiments tracking individual trajectories of highly Brownian rods under shear.

With the advent of improved imaging techniques such experiments are now possible and here we describe observations of the individual trajectories of rods which are about 3 microns in length as they move through microchannels under Poiseuille flow. Orbits are modified substantially in the presence of strong thermal noise which affects the orientational motion of individual rods.
In particular we find that, for rods a few microns in length,
Jeffery orbits can only be observed at very large shear rates, and that
the persistence of Jeffery orbits is very sensitive to the size of the rods. Reconstructing the full three-dimensional distribution of orientations we quantify the competition between Jeffery rotation and strong rotational Brownian motion which leads to fast decorrelation of the orientational state of a rod.  We quantify this in terms of the decay of the temporal autocorrelation of the Jeffery `constant': whereas 30  $\mu$m rods are well described by neglecting rotational diffusion \cite{Einarsson}, rods of 3.3  $\mu$m typically do not complete even a single complete Jeffery rotation. Using Brownian Dynamics simulations we show that Brownian noise, Jeffery rotation, and gravity are sufficient to explain the experimental results.

Silica rods with length 
$3-4\mu m$,
 diameter $0.5-0.7\mu m$ and average aspect ratio $\lambda=5.5$ \cite{Gao2015} were dispersed in deionised water.  The concentration of rods was  10 particles per $nl$ at most, such that the particles were far apart and did not interact with each other. The rods were not perfectly symmetric but bullet-shaped, with one end being a spherical cap while the other end was straight \cite{Kuijk2011}. 
%\juc{I've put everything in the past tense for consistency}
Microfluidic devices with a simple, unbranched channel, of height $H=10\mu m$ and width $300\mu m$, were prepared using standard soft-lithography techniques for PDMS. The channel consisted of two bends which were purely to increase the length of the channel, so that the rods had enough time to settle and
reach a steady state before being imaged.  Particles were imaged in the centre of the second straight segment, such that the effect of the bends as well as the lateral side walls  could be neglected.
Channels were plasma cleaned before use and a flow was achieved 
by imposing  a
 pressure difference between the inlet and outlet tubes. As the silica rods sediment easily,  the rods were expected to be distributed around  the bottom wall of the channel.

%where the \az{sign of the shear rate is constant}

%changes linearly with $z$ \juc{true everywhere if Poiseuille?}.  \juc{maybe need scaling arguments to estimate degree of sedimentation}

The system was imaged in the $xy$-plane, using a Zeiss LSM 5 Exciter in bright field mode, with a 63 x 1.4 NA oil-immersion objective. 
Image series of 1000 frames with a rate of $10$ frames per second were recorded using a Ximea MQ042MG-CM CMOS camera. Particles were thus tracked through time and their positions, orientations in the $xy$-plane, $\phi=\arctan(n_y/|n_x|)$, and lengths $L_p$ projected to the $xy$ plane determined in each frame using image analysis techniques. 
%\juc{MOVE?$L_p$ was used to determine  the $z$-component of the orientation \az{$n_z$} (see below), which is therefore always positive.
%\az{Note that the sign of $n_z$ does not influence the Jeffery orbit constant $C$.} }
In total we analyzed the data for 115 rods which were tracked for at least for 25 frames or 2.5$s$.

\begin{figure}
%	\centering
        \includegraphics[width=\columnwidth]{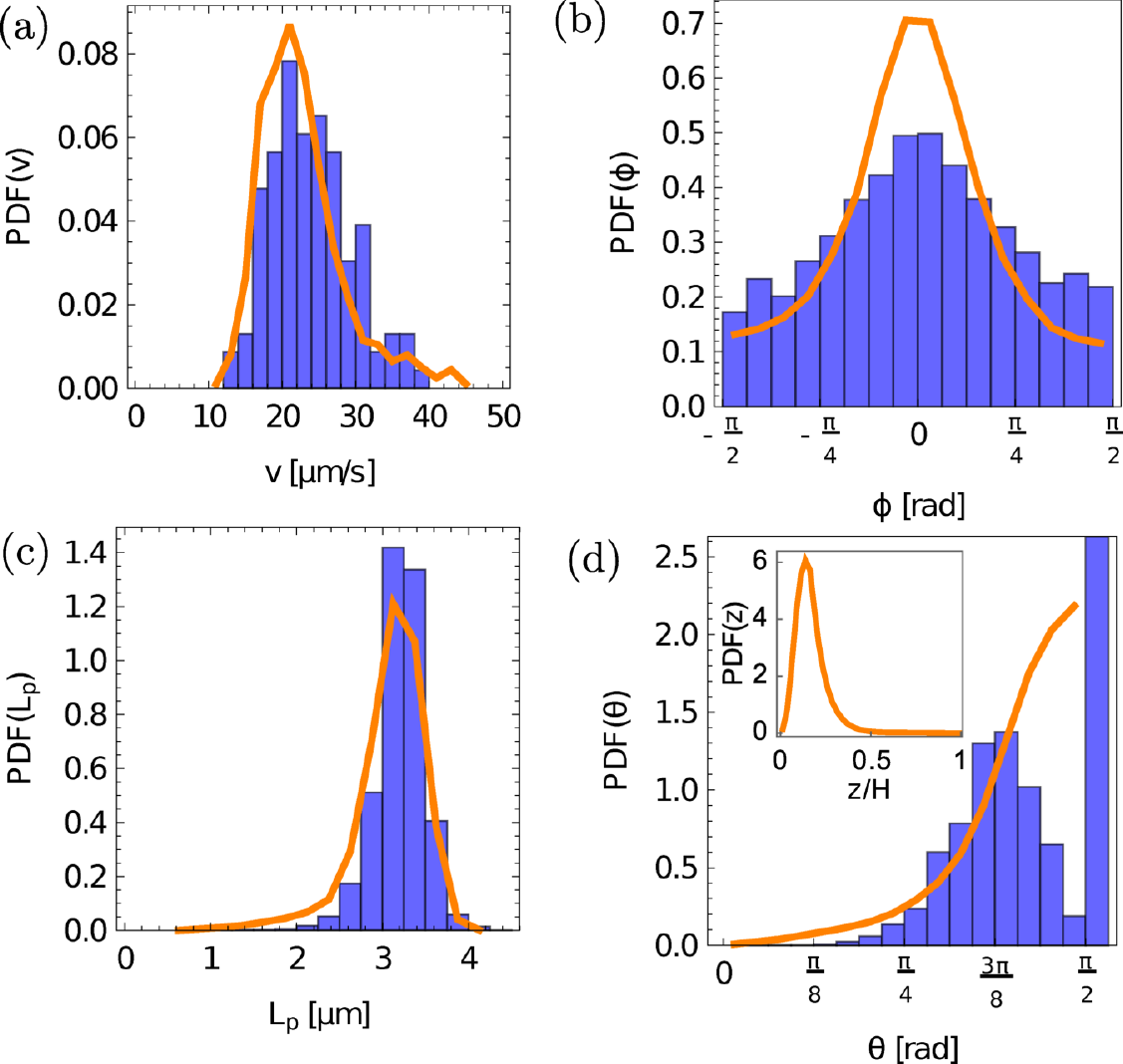}
	\caption{
Rod distribution functions for (a) velocities, (b) measured in-plane orientations $\phi$, (c) measured projected length $L_p$, (d)
calculated out-of-plane angles $\theta$ and lateral positions $z$ in the channel (inset).
The blue bars are experimental distributions and the orange lines are obtained from Brownian dynamics simulations.
\label{Fig:2}}
\end{figure}

The instantaneous and average velocity of each particle was determined by calculating the distance travelled in a frame or complete track respectively, and dividing by the time. The instantaneous velocity was found to be relatively constant for all particles, changing by no more than a few $\mu m s^{-1}$ throughout the time the particle was tracked. In contrast, the mean velocities of individual rods are spread in a  range of $10-40\mu m s^{-1}$ and peaked around $25\mu m s^{-1}$, see Fig.~\ref{Fig:2}(a). As the width and length of the channels are much larger than their height, the flow profile was comparable to a Hele-Shaw flow \cite{Tabeling}, i.e.\ a plane Poiseuille flow in the $xz$-plane, whereas the flow velocity was constant in the $xy$-plane except close to the side walls. 
The particles flowing through the channels therefore experienced a non-uniform, asymmetric shear in the $xz$-plane, see  Fig.~\ref{Fig:1}(a).
%As the silica rods  sediment easily, rods were typically near the bottom wall of the 
%channel, in agreement with the theoretical analysis presented below.
Since in Poiseuille flow the velocity of rods is connected to the lateral position in the channel, we conclude that different rods move at different average distances from the bottom surface but stay at almost constant height for individual tracks.

%The flow velocity can have an influence on the orientational behavior of a rod in two ways. Firstly, a higher flow velocity results in a stronger and less symmetric shear force experienced by the particle at the same $z$-position. Secondly, at a high flow velocity, a rod-shaped particle will tend to align with the flow-direction to minimise drag \cite{Probstein}.
%\juc{need to write in context of Jeffery orbits}

 Examples of rod behavior, illustrated in Fig.~\ref{Fig:1}, show clearly that the orientational behaviour of the rods did not follow simple Jeffery orbits.  %\juc{strange not to mention ordinary tumbling}
For some rods Jeffery-like kayaking behavior was observed  (Fig.~\ref{Fig:1}(c)) where the rod's orientation $n_x$  occasionally  flipped by approximately $180^\circ$ while $n_y$ 
%(almost always)
 remained either positive or negative (Fig.~\ref{Fig:1}(d)).
However, most of the rods were able to change the sign of $n_y$ dynamically so that the rods then tumbled in the $xy$ plane (see Fig.~\ref{Fig:1}(b,e)), which is not possible for non-Brownian Jeffery orbits.
Other trajectories showed even more irregular motion where no clear oscillation could be identified [see for example Fig.~\ref{Fig:1}(f)].
Simple Jeffery tumbling motion in the $xz$ plane was not observed.
%\juc{hard to see what is going on in (b),(c). - I'd show Jeffrey tunbling, Jeffrey kayaking, xy tumbling, random}

 Figure \ref{Fig:2}(b) shows the time-and ensemble averaged distribution for the angle $\phi$ which characterises the rod orientation in the $xy$-plane. % \juc{repeats but maybe clearer}.
The distribution is peaked and almost symmetric around zero, meaning the particles are most likely to align with the flow-direction. 
The lengths $L_p$ of the rods projected to the $xy$ plane are plotted in Fig.~\ref{Fig:2}(c). 
%\juc{DISCUSS, THE DISTRIBUTION LOOKS QUITE NARROW TO ME}
 The distribution peaks around $3.1\mu m$ and does not have a sharp cut-off to larger lengths due to the polydispersity of the rods and due to effects coming from the finite resolution of the image analysis. %\juc{what effects?}
The width of the distribution to smaller lengths stems from the out-of-plane orientation of rods.

%\juc{this paragraph needs rewriting, if we want to keep this data in, starting from something like The angle $\theta$ - not sure it is defined anywhere - is difficult to measure because ... An estimate of its distribution can be obtained from ...}
While the angle $\phi$ can be extracted directly from the tracked rods in the $xy$-plane, determining the out-of-plane angle $\theta=\arcsin |n_z|$ 
is less obvious.
The length of the rod in the $z$ direction is unknown and cannot be simply extracted from $L_p$
 due to the small polydispersity in the length $L$ of our rods.
% \juc{[MOVED, OMIT? $L_p$ was used to determine  the $z$-component of the orientation $n_z$ (see below), which is therefore always positive.
%Note that the sign of $n_z$ does not influence the Jeffery orbit constant $C$.]} 
%The tail to the left hints to the particle orientations $\theta$ and
Following Ref.~\cite{Gao2015}, $\theta$ 
is calculated using $\theta=\arcsin\left(\frac{\lambda L_p/L-1}{\lambda-1}\right)$.
%where $L$ is the length of an individual rod.
Since it is not possible to measure the length $L$ of each individual rod, we use
$L=3.3\mu m$, the average rod length, for all rods.
Hence,  $\theta$ can only be estimated, and it is possible that $L_p>L$.
In this case
 we always assume that the rod is aligned in the $xy$-plane, where $\theta=\pi/2$.
 In Fig.~\ref{Fig:2}(d) we show the distribution of $\theta$ which includes the artificial peak at $\theta= \pi/2$.
 Note that while we can fully determine the  orientation components $n_x$ and $n_y$, which lie in the range ($-1,1$), we can estimate $n_z=\sin\theta$ only up to a sign,
 and we chose $n_z>0$, see Fig.~\ref{Fig:1}.
 
% coming from the unknown length of the rod and the uncertainty of the image analysis process.

The observed orientational behavior of the silica rods deviates in many ways from Jeffery's theory. The most obvious difference lies in the aperiodicity of the dynamics  and the apparent random jumping between orbits characterised by different values of the Jeffery constant $C$. 
%This type of motion has been predicted theoretically for triaxial particles \cite{Yarin1997} and has been observed for non-Brownian rods \cite{einarsson}, where the authors again \ju{attributed the} effect to the slight asymmetry of their particles. It is possible that the asymmetry of the ends of our silica rods leads to the seemingly random behavior observed. 
%Another possible explanation is that 
The Brownian motion of the rods changes their orientation in such a way that they move continually from one orbit to another, leading to randomness in the orientational dynamics. %\juc{implied question here that we don't answer}

In order to identify the relative importance and interplay of shear rate, Brownian noise, gravity and confinement, we performed Brownian Dynamics (BD) simulations of  an ellipsoid in Poiseuille flow using the experimental parameters. 
%\juc{[need to flag up the comparison to simulations in the intro, and also make sure these factors are mentioned -- they are but not together in a way that asks this question earlier]} \az{[yes, maybe still have to be improved in intro]}
A rod is approximated by a prolate ellipsoid of length $L=3.3 \mu m$, width $W=0.6 \mu m$,  aspect ratio $\lambda=5.5$ and volume $V = 0.62 \mu m^3$.
We approximate the parallel and perpendicular translational ($D_{||}$ and $D_{\perp}$) and rotational ($D_r$) diffusion coefficients of a rod at room temperature and water viscosity using analytic expressions for a prolate ellipsoid \cite{KimKarila}, which gives $D_{||}=0.39\mu m^2s^{-1}$,  $D_{\perp}=0.29\mu m^2s^{-1}$, $D_r=0.21s^{-1}$.
Since the rods are approximately 90\% heavier than water, they experience a density difference  $\Delta \rho = 900 kg/m^3$ \cite{Gao2015}
%The density different $\Delta \rho$ between the rods and water is approximately 90\% %\juc{not sure about the definiton of $\Delta \rho$}
resulting in a typical sedimentation velocity $v_s = V \Delta \rho g/\gamma_s = 0.46 \mu m s^{-1}$, where $\gamma_s$ is a typical friction coefficient estimated as $\gamma_s=2k_BT/(D_{||} + D_\perp)$.
% \az{[exact definition is not so important since $D_{||}$ similar to $D_\perp$]}.
The corresponding gravitational P\'eclet number  \cite{Enculescu2011} 
$\alpha=V \Delta \rho g L/ k_BT =  4.2$, or, equivalently, the sedimentation length is $L_\text{sed} =  L/ \alpha = 0.7\mu m$.
Hence $L_\text{sed} + L/2$
  can be interpreted as the typical length a rod is away from the bottom wall due to thermal fluctuations and steric effects. % \juc{??in addition to steric repulsion}. 
%\az{[We can shorten this if we need space]}

The rod moves in a planar Poiseuille flow $\mathbf{v}_f$ characterized by a wall shear rate $\dot{\gamma}$. We use the experimental channel dimensions  and do not allow the rod to penetrate the walls.
Brownian dynamics simulations are used to calculate the rod position $\mathbf{r}$ and orientation $\mathbf{n}$:
\begin{equation}
\dot{\mathbf{r}} = \mathbf{v}_f - v_s\mathbf{\hat{z}} + \mathcal{H} \cdot \boldsymbol{\xi}, \quad \dot{\mathbf{n}} = \boldsymbol{\Omega}_J(\mathbf{n};z) + \sqrt{2D_r}\boldsymbol{\xi}_r \times \mathbf{n}
\end{equation}
where $\mathcal{H}$ accounts for the translational diffusion of the rod (for details see Ref.~\cite{Matsunaga2017}), $\boldsymbol{\Omega}_J(\mathbf{n};z)$
 is Jeffery's reorientation rate which depends linearly on the local shear rate $\dot{\gamma}_l(z)=\dot{\gamma}(1-2z/H)$, and $\boldsymbol{\xi}$ and $\boldsymbol{\xi}_r$ denote Gaussian white noise.
 
We perform simulations for different wall shear rates $\dot{\gamma}$, averaging over 5000 rods  with random initial conditions for each value of the shear. 
%\juc{averaging is for how long, and mention equilibration time?} 
We find, as expected, that 
after equilibration
all the rods are distributed in the lower half of the channel because of sedimentation due to gravity, see inset of Fig.~\ref{Fig:2}(d).
The  shear rate $\dot{\gamma}$
%We do not know the shear rate $\dot{\gamma}$ and particle positions $z$ from the experiments. However these 
could be inferred from measurements of the distribution of particle velocities:
using $\dot{\gamma}=18s^{-1}$ results in a velocity distribution that matches the experiments very well (see the orange curve in Fig.~\ref{Fig:2}(a)). This value is used to create all of the simulation results shown in Figs.~2 and 3.
We note that without including the sedimentation of the rods the velocity distribution cannot be reproduced. % \ju{\sout{(not shown)}}.

We identify the same qualitative orientational behavior as observed in the experiments, as shown by the examples in Fig.~\ref{Fig:1}(d-f). As a more quantitative comparison we
compare the orientation distributions for the angles $\phi$ and $\theta$ in Fig.~\ref{Fig:2}(b,d), obtaining the same trends as in the experiments.
The peak for $\phi$ is  more pronounced in the simulations which could result from  the small shape asymmetry of the rods and possible hydrodynamic rod-wall interactions  which are neglected in the simulations.
The distribution for $\theta$ essentially captures the experimental values, but  does not show the artificial sharp peak at $\theta=\pi/2$.
We also compare the distributions of projected rod lengths $L_p$ in Fig.~\ref{Fig:2}(c). The simulations reproduce the experiments well 
if we assume that the rod length $L$ is normally distributed with standard deviation $\sigma = 0.25 \mu m$.
%\az{The distributions of the $y$ position in the channel is shown in the inset of Fig.~\ref{Fig:2}(d) confirming that all rods are located at the lower half of the channel. }

Brownian motion gives an explanation for the $xy$-tumbling motion, which is not predicted by Jeffery theory, and which has not been observed for anisotropic particles.  If the rods follow a kayaking trajectory with a large amplitude, they come very close to aligning with the $x$-axis after each flip. In a Jeffery orbit a rod spends most of its time in this flow-aligned orientation. Similarly here the particle will spend some time performing  Brownian motion around the flow-aligned direction, which could pull it to the other side of the $x$-axis before it starts another kayaking cycle, effectively making it tumble in the $xy$-plane. 
This is confirmed by our simulations, and an example trajectory is shown in Fig.~\ref{Fig:1}(e).

\begin{figure}
%  \centering
  \includegraphics[width=\columnwidth]{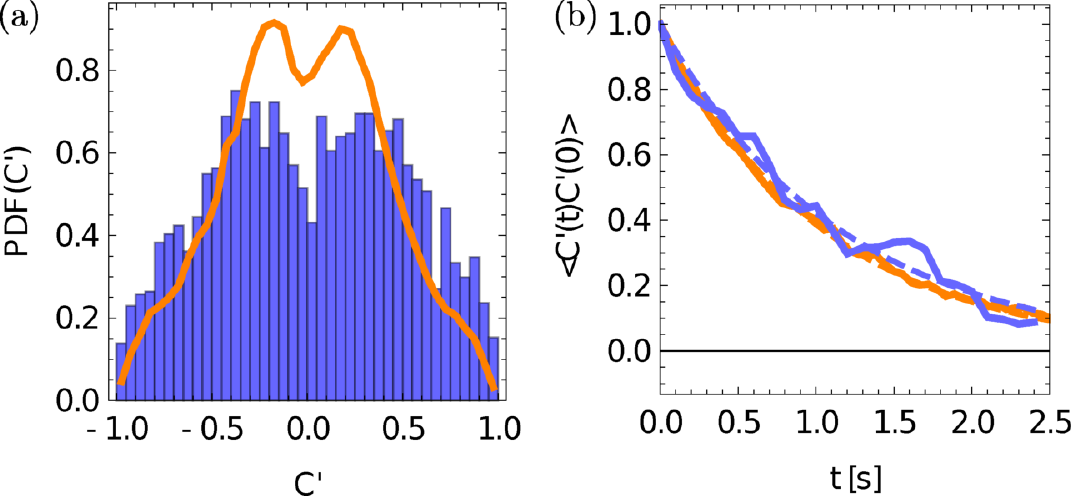}
  \caption{
(a) Distribution of modified Jeffery  constants $C'$.
Color code as in Fig.~2.
 (b) Temporal auto-correlation function of $C'$ including exponential fits (blue: experiments; orange: simulations).
}
%The blue bars are experimental distributions and the orange lines are obtained from Brownain dynamics simulations.}
 \label{Fig:3}
\end{figure}

To quantify the competition between periodic Jeffery orbits and the influence of noise we determine the instantaneous Jeffery  ``constants'' $C(t)$ from  both experiments and simulations.
%\az{Note that $C(t)$ only depends on the absolute value of $n_z$ and is hence  well-defined for the experimental  }
For convenience we use a modified Jeffery  constant $C'=\sign(C)/(1+|C|)$ which maps to the interval $C' \in \{-1,1 \}$ where $C'=0$ corresponds to $C=\pm \infty$ (rotation in $xz$ plane) and $C'=\pm 1$ to $C=0$ (log rolling).
The particle- and time-averaged distributions of $C'$, which are shown in Fig.~\ref{Fig:3}(a),
match reasonably well between experiments and simulations. 
In particular they show maxima in the distributions around $C' \sim \pm 0.25$.
To determine the persistence of a Jeffery orbit, we compute the temporal $C'$ auto-correlation function
$\langle C'(t)C'(0) \rangle$ which decays approximately exponentially as $~\exp(-t/ \tau)$ (Fig.~\ref{Fig:3}(b)).
The decay time $\tau\approx 1s$ agrees very well between experiments ($\tau=1.16s$) and simulations ($\tau=1.05s$).
% This shows that a rod loses information about its Jeffery orbit state. 

It is instructive to compare $\tau$ to the Jeffery oscillation period $t_J=2\pi(\lambda+\lambda^{-1})/\dot{\gamma} = 1.98s$.
$\tau/t_J$ gives the number of Jeffery oscillations a rod performs before losing information about its Jeffery orbit state.
%, i.e.\ when characterizing the rod orientation simply by its orbit constant starts to fail .
Since in our system $\tau/t_J\approx 0.5$ the rod does not even finish one turn before its state decorrelates,  reiterating that we do not observe clear Jeffery orbits in the experiments or simulations.

\begin{figure}[htb]
        \includegraphics[width=\columnwidth]{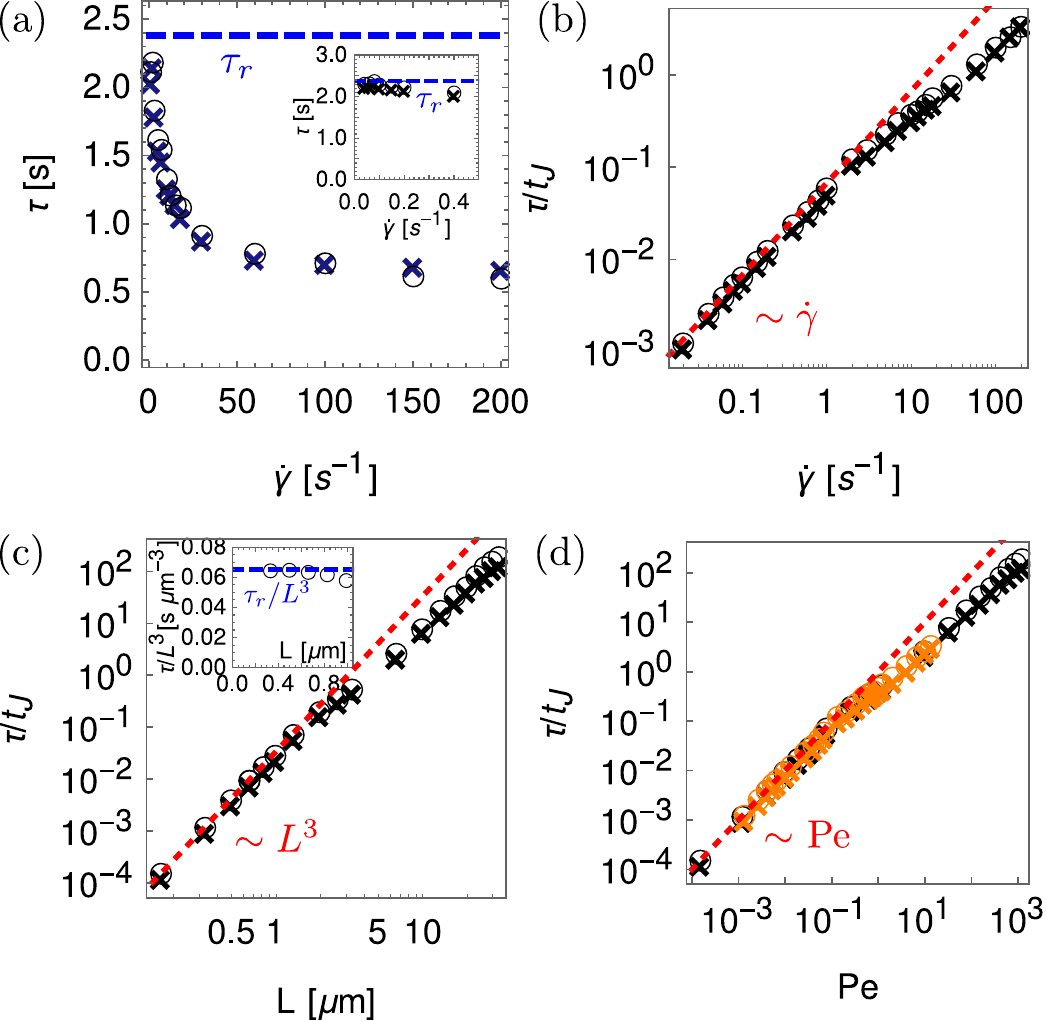}
	\caption{
          (a) Dependence of the rods' Jeffery orbit persistence time $\tau$  on the shear rate $\dot{\gamma}$. The blue dashed line shows the rotational diffusion time $\tau_r$.
            Inset: For small $\dot{\gamma}$ the decay time $\tau$ becomes comparable to $\tau_r$.
            (b) $\tau$ normalized by the Jeffery period $t_J$ as a function of  $\dot{\gamma}$.
            The red dashed line shows the linear behavior for small shear rates $\tau_r/ \tau_J\sim \dot{\gamma}$.
            (c) Dependence of $\tau/t_J$ on the length $L$ of the rods for a rod aspect ratio $\lambda=5.5$ and shear rate $\dot{\gamma}=18s^{-1}$.  The red dashed line shows the  behavior for small rod length $\tau_r/ \tau_J \sim L^3$. Inset: $\tau$ approaches $\tau_r$ for rod lengths <$\sim$  half a micron.
            (d)  Dependence of $\tau/t_J$ on the rotational P\'eclet number Pe. The red dashed line shows the linear behavior for small Pe.
            Black symbols show the data set for varying rod length, while orange symbols show the data set for varying shear rate.
  In (a-d) crosses indicate full 3D simulations, and circles simulations for rods fixed at position $z=0.2H$. % \az{[check symbol definition]}.
\label{Fig:4}}
\end{figure}

%\ju{unflagged changes in this paragraph\\}
In Fig.~\ref{Fig:4}(a) we show the dependence of the Jeffery decorrelation time $\tau$  on the wall shear rate $\dot{\gamma}$.
  We observe that $\tau$ decreases with increasing $\dot{\gamma}$ meaning that stronger shear leads to faster decorrelation.
%  For sufficiently large $\dot{\gamma}$ the decay of $\tau$ can be fitted to a power law $\tau \sim \dot{\gamma}^{-0.2}$.
%  fits well to a power law $\tau \sim \dot{\gamma}^{-\frac{1}{4}}$ MODIFY}.
Interestingly, $\tau$ is  smaller than the rotational diffusion time $\tau_r=1/(2D_r)=2.38s$ [dashed blue curve in  Fig.~\ref{Fig:4}(a)] but  approaches $\tau_r$ for vanishing shear rates [see inset of Fig.~\ref{Fig:4}(a)], when the system is essentially governed by Brownian fluctuations only. 
Thus, at higher rates there is a competition between the relatively fast rotation on the unit sphere helping the fluctuations to faster decorrelate the memory of the Jeffery orbits and faster shear rates, and hence rotation rates, helping rods to finish Jeffery rotations before they are fully decorrelated. This can be seen in Fig.~\ref{Fig:4}(b), where we plot $\tau $ per Jeffery reorientation time $t_J$ which increases 
%The inset shows that 
%the number of persistent Jeffery oscillations behave as
%$\tau/t_J \sim \dot{\gamma}^{0.8}$ and
linearly with $\dot{\gamma}$ for small $\dot{\gamma}$, and sub-linearly for higher $\dot{\gamma}$, in accordance with  Fig.~\ref{Fig:4}(a),  since $t_J \sim \dot{\gamma}^{-1}$. This    indicates that only very large  shear rates of $\mathcal{O}(10^2s^{-1})$  would allow  clear Jeffery orbits to be observed for several oscillations in experiments using our rods. %, \juc{?similar as for tumbling bacterial cell bodies in flow \cite{Kaya2009}.}
Indeed such clear orbits have been observed for tumbling bacterial cell bodies in strong shear \cite{Kaya2009}.

We also analyze the dependence of $\tau/t_J$ on the rod length $L$  while keeping the shear rate $\dot{\gamma}=18s^{-1}$ and $L/H=0.33$ constant.
Fig.~\ref{Fig:4}(c) shows   a strong dependence on the rod length. For example, $10\mu m$ long rods would already perform about 10 persistent oscillations.
  Again, for very small lengths $L$ the orientational dynamics is mainly governed by rotational Brownian noise since $\tau_r \sim L^3$.
  Indeed we show in the inset of Fig.~\ref{Fig:4}(c) that $\tau \sim L^3 $ and that $\tau$ approaches $\tau_r$ for rod lengths on the order of $L \lesssim 500nm$.

  The fact that for both small $\dot{\gamma}$ and small $L$ the dynamics is governed by Brownian fluctuations, while for large $\dot{\gamma}$ and $L$ the dynamics becomes determinstic,
  can be captured by the rotational P\'eclet number Pe=$\tau_r/t_J= f(\lambda)\dot{\gamma}/D_r$ with $f(\lambda)=[4\pi(\lambda+\lambda^{-1})]^{-1}$ which compares the rotational diffusion time with the Jeffery reorientation time.
  For our rods the shape function $f(\lambda)=f(5.5)=0.014$ and our experiments are performed at $\text{Pe}=1.2$. (Note that a common alternative definition of Pe omits this shape function.)
 
  We plot the results for both varying shear rate and varying rod length, as discussed in Fig.~\ref{Fig:4}(a-c), as a function of Pe in Fig.~\ref{Fig:4}(d).
  Indeed the data for varying $\dot{\gamma}$ (shown in orange) and varying $L$ (shown in black) collapses to a single curve.
%  and again the dependence of $\tau$ on Pe can be fitted to a power law $\tau\sim \text{Pe}^{0.8}\sim(\dot{\gamma}L^3)^{0.8}\sim \dot{\gamma}^{0.8}L^{2.4}$ for sufficiently large Pe, which is in reasonable agreement with the exponents extracted from the two individual data sets.
  As expected, for very small Pe, $\tau\sim Pe$.

Finally we note that these results are  not a consequence of gravity- and noise-induced cross-streamline migration or the steric interaction with the walls:
Keeping the $z$ position of the particles fixed, for example at $z=0.2H$, does not modify the results significantly, as shown in Fig.~\ref{Fig:4}.

%\az{What is Jeffery frequency?}

%This is quantified by a change of sign in $C$ or $C'$.

%ADD SIMULATION RESULTS TO FIG 1c TO SHOW THAT XY TUMBLING ONLY COMES FROM NOISE? \juc{would be nice; maybe to fig 3}

%SHOW HOW $tau$ DEPEND ON PARAMETERS (PARTICLE SIZE< SHEAR RATE ETC) WITH SIMULATIONS?  \juc{a parameter scan cold be a nice conference paper}

%\juc{will need extending later}
We have studied colloidal rods flowing in a plane Poiseuille flow, and observed particles performing stochastic kayaking and $xy$-tumbling motions. The latter behavior is not modelled within Jeffery theory, but can be explained by the Brownian nature of the rods. 
Rods are able to switch between different states,  and their dynamics can be quantified by the Jeffery  constant and its temporal correlations.
Based on our findings, it would be interesting to study in the future the influence of the Brownian motion on denser suspensions of colloidal rods.

\section*{Conflicts of interest}
There are no conflicts to declare.

\section*{Acknowledgements}
This work was supported through funding from the ERC Advanced Grant 291234 MiCE.
A.Z. acknowledges funding
 from the Austrian Science Fund (FWF) through a Lise-Meitner Fellowship (Grant No M 2458-N36),
and from the European Union
through a Marie Sk{\l}odowska Curie
Intra-European Fellowship (G.A. No 653284).
 Y.G. acknowledges financial support from the General Program of the National Natural Science Foundation of China (Project no.11774237).

%%%END OF MAIN TEXT%%%

 %For footnotes in the main text of the article please number the footnotes to avoid duplicate symbols. e.g.  \footnote[num]{your text} the corresponding author \ast counts as footnote 1, ESI as footnote 2, e.g. if there is no ESI, please start at [num]=[2], if ESI is cited in the title please start at [num]=[3] etc. Please also cite the ESI within the main body of the text using \dag.

%The \balance command can be used to balance the columns on the final page if desired. It should be placed anywhere within the first column of the last page.

%\balance

%If notes are included in your references you can change the title from 'References' to 'Notes and references' using the following command:
%\renewcommand\refname{Notes and references}

%%%REFERENCES%%%
\scriptsize{
\bibliography{ch8} %You need to replace "rsc" on this line with the name of your .bib file

\providecommand*{\mcitethebibliography}{\thebibliography}
\csname @ifundefined\endcsname{endmcitethebibliography}
{\let\endmcitethebibliography\endthebibliography}{}
\begin{mcitethebibliography}{29}
\providecommand*{\natexlab}[1]{#1}
\providecommand*{\mciteSetBstSublistMode}[1]{}
\providecommand*{\mciteSetBstMaxWidthForm}[2]{}
\providecommand*{\mciteBstWouldAddEndPuncttrue}
  {\def\EndOfBibitem{\unskip.}}
\providecommand*{\mciteBstWouldAddEndPunctfalse}
  {\let\EndOfBibitem\relax}
\providecommand*{\mciteSetBstMidEndSepPunct}[3]{}
\providecommand*{\mciteSetBstSublistLabelBeginEnd}[3]{}
\providecommand*{\EndOfBibitem}{}
\mciteSetBstSublistMode{f}
\mciteSetBstMaxWidthForm{subitem}
{(\emph{\alph{mcitesubitemcount}})}
\mciteSetBstSublistLabelBeginEnd{\mcitemaxwidthsubitemform\space}
{\relax}{\relax}

\bibitem[Blanco \emph{et~al.}(2015)Blanco, Shen, and Ferrari]{BS15}
E.~Blanco, H.~Shen and M.~Ferrari, \emph{Nature Biotechnology}, 2015,
  \textbf{33}, 941\relax
\mciteBstWouldAddEndPuncttrue
\mciteSetBstMidEndSepPunct{\mcitedefaultmidpunct}
{\mcitedefaultendpunct}{\mcitedefaultseppunct}\relax
\EndOfBibitem
\bibitem[Junot \emph{et~al.}(2019)Junot, Figueroa-Morales, Darnige, Lindner,
  Soto, Auradou, and Cl{\'e}ment]{JF19}
G.~Junot, N.~Figueroa-Morales, T.~Darnige, A.~Lindner, R.~Soto, H.~Auradou and
  E.~Cl{\'e}ment, \emph{arXiv:1903.02995}, 2019\relax
\mciteBstWouldAddEndPuncttrue
\mciteSetBstMidEndSepPunct{\mcitedefaultmidpunct}
{\mcitedefaultendpunct}{\mcitedefaultseppunct}\relax
\EndOfBibitem
\bibitem[Fry \emph{et~al.}(2005)Fry, Langhorst, Kim, Grulke, Wang, and
  Hobbie]{FL05}
D.~Fry, B.~Langhorst, H.~Kim, E.~Grulke, H.~Wang and E.~K. Hobbie, \emph{Phys.
  Rev. Lett.}, 2005, \textbf{95}, 038304\relax
\mciteBstWouldAddEndPuncttrue
\mciteSetBstMidEndSepPunct{\mcitedefaultmidpunct}
{\mcitedefaultendpunct}{\mcitedefaultseppunct}\relax
\EndOfBibitem
\bibitem[Dickinson(2015)]{D15}
E.~Dickinson, \emph{Annual Review of Food Sceince and Technology}, 2015,
  \textbf{6}, year\relax
\mciteBstWouldAddEndPuncttrue
\mciteSetBstMidEndSepPunct{\mcitedefaultmidpunct}
{\mcitedefaultendpunct}{\mcitedefaultseppunct}\relax
\EndOfBibitem
\bibitem[Jeffery(1922)]{Jeffery}
G.~B. Jeffery, \emph{Proceedings of the Royal Society of London A:
  Mathematical, Physical and Engineering Sciences}, 1922, \textbf{102},
  161--179\relax
\mciteBstWouldAddEndPuncttrue
\mciteSetBstMidEndSepPunct{\mcitedefaultmidpunct}
{\mcitedefaultendpunct}{\mcitedefaultseppunct}\relax
\EndOfBibitem
\bibitem[Einarsson \emph{et~al.}(2015)Einarsson, Candelier, Lundell, Angilella,
  and Mehlig]{logrolling}
J.~Einarsson, F.~Candelier, F.~Lundell, J.~R. Angilella and B.~Mehlig,
  \emph{Phys. Rev. E}, 2015, \textbf{91}, 041002\relax
\mciteBstWouldAddEndPuncttrue
\mciteSetBstMidEndSepPunct{\mcitedefaultmidpunct}
{\mcitedefaultendpunct}{\mcitedefaultseppunct}\relax
\EndOfBibitem
\bibitem[Einarsson \emph{et~al.}(2013)Einarsson, Johansson, Mahato, Mishra,
  Angilella, Hanstorp, and Mehlig]{Einarsson}
J.~Einarsson, A.~Johansson, S.~K. Mahato, Y.~N. Mishra, J.~R. Angilella,
  D.~Hanstorp and B.~Mehlig, \emph{Acta Mechanica}, 2013, \textbf{224},
  2281--2289\relax
\mciteBstWouldAddEndPuncttrue
\mciteSetBstMidEndSepPunct{\mcitedefaultmidpunct}
{\mcitedefaultendpunct}{\mcitedefaultseppunct}\relax
\EndOfBibitem
\bibitem[Hinch and Leal(1979)]{HinchLeal}
E.~J. Hinch and L.~G. Leal, \emph{J. Fluid Mech.}, 1979, \textbf{92},
  591--607\relax
\mciteBstWouldAddEndPuncttrue
\mciteSetBstMidEndSepPunct{\mcitedefaultmidpunct}
{\mcitedefaultendpunct}{\mcitedefaultseppunct}\relax
\EndOfBibitem
\bibitem[Yarin \emph{et~al.}(1997)Yarin, Gottlieb, and Roisman]{Yarin1997}
A.~L. Yarin, O.~Gottlieb and I.~V. Roisman, \emph{J. Fluid Mech.}, 1997,
  \textbf{340}, 83--100\relax
\mciteBstWouldAddEndPuncttrue
\mciteSetBstMidEndSepPunct{\mcitedefaultmidpunct}
{\mcitedefaultendpunct}{\mcitedefaultseppunct}\relax
\EndOfBibitem
\bibitem[Einarsson \emph{et~al.}(2016)Einarsson, Mihiretie, Laas, Ankardal,
  Angilella, Hanstorp, and Mehlig]{Mehlig2016}
J.~Einarsson, B.~M. Mihiretie, A.~Laas, S.~Ankardal, J.~R. Angilella,
  D.~Hanstorp and B.~Mehlig, \emph{Phys. Fluids}, 2016, \textbf{28},
  013302\relax
\mciteBstWouldAddEndPuncttrue
\mciteSetBstMidEndSepPunct{\mcitedefaultmidpunct}
{\mcitedefaultendpunct}{\mcitedefaultseppunct}\relax
\EndOfBibitem
\bibitem[Hijazi and Khater(2001)]{HZ01}
A.~Hijazi and A.~Khater, \emph{Computational Materials Science}, 2001,
  \textbf{22}, 279\relax
\mciteBstWouldAddEndPuncttrue
\mciteSetBstMidEndSepPunct{\mcitedefaultmidpunct}
{\mcitedefaultendpunct}{\mcitedefaultseppunct}\relax
\EndOfBibitem
\bibitem[Zurita-Gotor \emph{et~al.}(2007)Zurita-Gotor, B{\l}awzdziewicz, and
  Wajnryb]{ZB07}
M.~Zurita-Gotor, J.~B{\l}awzdziewicz and E.~Wajnryb, \emph{Journal of
  Rheology}, 2007, \textbf{51}, 71\relax
\mciteBstWouldAddEndPuncttrue
\mciteSetBstMidEndSepPunct{\mcitedefaultmidpunct}
{\mcitedefaultendpunct}{\mcitedefaultseppunct}\relax
\EndOfBibitem
\bibitem[Atwi \emph{et~al.}(2013)Atwi, Khater, and Hijazi]{AK13}
A.~Atwi, A.~Khater and A.~Hijazi, \emph{Polymer}, 2013, \textbf{54}, 1555\relax
\mciteBstWouldAddEndPuncttrue
\mciteSetBstMidEndSepPunct{\mcitedefaultmidpunct}
{\mcitedefaultendpunct}{\mcitedefaultseppunct}\relax
\EndOfBibitem
\bibitem[Holmstedt \emph{et~al.}(2016)Holmstedt, Akerstedt, Lundstrom, and
  Hogberg]{HA16}
E.~Holmstedt, H.~O. Akerstedt, T.~S. Lundstrom and S.~M. Hogberg, \emph{J.
  Fluids Engineering}, 2016, \textbf{138}, year\relax
\mciteBstWouldAddEndPuncttrue
\mciteSetBstMidEndSepPunct{\mcitedefaultmidpunct}
{\mcitedefaultendpunct}{\mcitedefaultseppunct}\relax
\EndOfBibitem
\bibitem[Monjezi \emph{et~al.}(2018)Monjezi, Jones, Nelson, and Park]{MJ18}
S.~Monjezi, J.~D. Jones, A.~K. Nelson and J.~Park, \emph{Nanomaterials}, 2018,
  \textbf{8}, 130\relax
\mciteBstWouldAddEndPuncttrue
\mciteSetBstMidEndSepPunct{\mcitedefaultmidpunct}
{\mcitedefaultendpunct}{\mcitedefaultseppunct}\relax
\EndOfBibitem
\bibitem[Gunes \emph{et~al.}(2008)Gunes, Scirocco, Mewis, and Vermant]{GS08}
D.~Z. Gunes, R.~Scirocco, J.~Mewis and J.~Vermant, \emph{J. Non-Newtonian Fluid
  Mechanics}, 2008, \textbf{155}, 39\relax
\mciteBstWouldAddEndPuncttrue
\mciteSetBstMidEndSepPunct{\mcitedefaultmidpunct}
{\mcitedefaultendpunct}{\mcitedefaultseppunct}\relax
\EndOfBibitem
\bibitem[Frattini and Fuller(1986)]{FF86}
P.~L. Frattini and G.~G. Fuller, \emph{J. Fluid Mech.}, 1986, \textbf{168},
  119\relax
\mciteBstWouldAddEndPuncttrue
\mciteSetBstMidEndSepPunct{\mcitedefaultmidpunct}
{\mcitedefaultendpunct}{\mcitedefaultseppunct}\relax
\EndOfBibitem
\bibitem[Xie \emph{et~al.}(2015)Xie, Lista, Qiao, and Dunstan]{XL15}
D.~Xie, M.~Lista, G.~G. Qiao and D.~E. Dunstan, \emph{J. Physical Chemistry
  Letters}, 2015, \textbf{6}, 3815\relax
\mciteBstWouldAddEndPuncttrue
\mciteSetBstMidEndSepPunct{\mcitedefaultmidpunct}
{\mcitedefaultendpunct}{\mcitedefaultseppunct}\relax
\EndOfBibitem
\bibitem[Hinch and Leal(1973)]{HL73}
E.~J. Hinch and L.~G. Leal, \emph{J. Fluid Mech.}, 1973, \textbf{57}, 753\relax
\mciteBstWouldAddEndPuncttrue
\mciteSetBstMidEndSepPunct{\mcitedefaultmidpunct}
{\mcitedefaultendpunct}{\mcitedefaultseppunct}\relax
\EndOfBibitem
\bibitem[Hijazi and Zoaeter(2002)]{HZ02}
A.~Hijazi and M.~Zoaeter, \emph{European Polymer J.}, 2002, \textbf{38},
  2207\relax
\mciteBstWouldAddEndPuncttrue
\mciteSetBstMidEndSepPunct{\mcitedefaultmidpunct}
{\mcitedefaultendpunct}{\mcitedefaultseppunct}\relax
\EndOfBibitem
\bibitem[Leahy \emph{et~al.}(2015)Leahy, Koch, and Cohen]{LK15}
B.~D. Leahy, D.~L. Koch and I.~Cohen, \emph{J. Fluid Mech.}, 2015,
  \textbf{772}, 42\relax
\mciteBstWouldAddEndPuncttrue
\mciteSetBstMidEndSepPunct{\mcitedefaultmidpunct}
{\mcitedefaultendpunct}{\mcitedefaultseppunct}\relax
\EndOfBibitem
\bibitem[Palanisamy and den Otter(2018)]{PO18}
D.~Palanisamy and W.~K. den Otter, \emph{J. Chem. Phys.}, 2018, \textbf{148},
  194112\relax
\mciteBstWouldAddEndPuncttrue
\mciteSetBstMidEndSepPunct{\mcitedefaultmidpunct}
{\mcitedefaultendpunct}{\mcitedefaultseppunct}\relax
\EndOfBibitem
\bibitem[Gao \emph{et~al.}(2015)Gao, Balin, Dullens, Yeomans, and
  Aarts]{Gao2015}
Y.~Gao, A.~K. Balin, R.~P. Dullens, J.~M. Yeomans and D.~G. Aarts, \emph{Phys.
  Rev. Lett.}, 2015, \textbf{115}, 248301\relax
\mciteBstWouldAddEndPuncttrue
\mciteSetBstMidEndSepPunct{\mcitedefaultmidpunct}
{\mcitedefaultendpunct}{\mcitedefaultseppunct}\relax
\EndOfBibitem
\bibitem[Kuijk \emph{et~al.}(2011)Kuijk, van Blaaderen, and Imhof]{Kuijk2011}
A.~Kuijk, A.~van Blaaderen and A.~Imhof, \emph{Journal of the American Chemical
  Society}, 2011, \textbf{133}, 2346--2349\relax
\mciteBstWouldAddEndPuncttrue
\mciteSetBstMidEndSepPunct{\mcitedefaultmidpunct}
{\mcitedefaultendpunct}{\mcitedefaultseppunct}\relax
\EndOfBibitem
\bibitem[Tabeling(2005)]{Tabeling}
P.~Tabeling, \emph{Introduction to Microfluidics}, Oxford University Press,
  Oxford, 2005\relax
\mciteBstWouldAddEndPuncttrue
\mciteSetBstMidEndSepPunct{\mcitedefaultmidpunct}
{\mcitedefaultendpunct}{\mcitedefaultseppunct}\relax
\EndOfBibitem
\bibitem[Kim and Karila(2005)]{KimKarila}
S.~Kim and S.~J. Karila, \emph{Microhydrodynamics: Principles and Selected
  Applications}, Dover Publications Inc., Mineaola, New York, 2005\relax
\mciteBstWouldAddEndPuncttrue
\mciteSetBstMidEndSepPunct{\mcitedefaultmidpunct}
{\mcitedefaultendpunct}{\mcitedefaultseppunct}\relax
\EndOfBibitem
\bibitem[Enculescu and Stark(2011)]{Enculescu2011}
M.~Enculescu and H.~Stark, \emph{Phys. Rev. Lett.}, 2011, \textbf{107},
  058301\relax
\mciteBstWouldAddEndPuncttrue
\mciteSetBstMidEndSepPunct{\mcitedefaultmidpunct}
{\mcitedefaultendpunct}{\mcitedefaultseppunct}\relax
\EndOfBibitem
\bibitem[Matsunaga \emph{et~al.}(2017)Matsunaga, Meng, Z\"ottl, Golestanian,
  and Yeomans]{Matsunaga2017}
D.~Matsunaga, F.~Meng, A.~Z\"ottl, R.~Golestanian and J.~M. Yeomans,
  \emph{Phys. Rev. Lett.}, 2017, \textbf{119}, 198002\relax
\mciteBstWouldAddEndPuncttrue
\mciteSetBstMidEndSepPunct{\mcitedefaultmidpunct}
{\mcitedefaultendpunct}{\mcitedefaultseppunct}\relax
\EndOfBibitem
\bibitem[Kaya and Koser(2009)]{Kaya2009}
T.~Kaya and H.~Koser, \emph{Phys. Rev. Lett.}, 2009, \textbf{103}, 138103\relax
\mciteBstWouldAddEndPuncttrue
\mciteSetBstMidEndSepPunct{\mcitedefaultmidpunct}
{\mcitedefaultendpunct}{\mcitedefaultseppunct}\relax
\EndOfBibitem
\end{mcitethebibliography}
\bibliographystyle{rsc} } %the RSC's .bst file

\end{document}